\documentclass[lineno]{jfm}

\usepackage{graphicx}
\usepackage{epstopdf,epsfig}
\usepackage{newtxtext}
\usepackage{newtxmath}
\usepackage{bm}
\usepackage{natbib}
\usepackage{hyperref}
\usepackage{amsmath}

\usepackage{tikz}
\usepackage{comment}
\usepackage{adjustbox}
\usepackage{float}
\usepackage{xcolor}
\usepackage{pgfplots}
\usepackage{tikz-3dplot}
\usepackage{pgfplots}
\usepackage{paralist}
\usepackage{array}
\usepackage{booktabs}
\usepackage{ragged2e}
\usepackage{soul}
\usepackage{orcidlink}
\pgfplotsset{compat=1.17}
\usepgfplotslibrary{groupplots}
\usepackage{CJKutf8}
\definecolor{orcidlogocol}{HTML}{A6CE39}
\hypersetup{
    colorlinks = true,
    urlcolor   = blue,
    citecolor  = black,
}

\newcommand{\RomanNumeralCaps}[1]
\linenumbers

\title{Actuation manifold from snapshot data}

\author{Luigi Marra\aff{1}\corresp{\email{luigi.marra@uc3m.es}}, Guy Y. Cornejo Maceda\aff{2}, Andrea Meil\'an-Vila\aff{3}, Vanesa Guerrero\aff{3}, Salma Rashwan\aff{1}, Bernd R. Noack\aff{2,4}\corresp{\email{bernd.noack@hit.edu.cn}}, Stefano Discetti\aff{1}, \and
  Andrea Ianiro\aff{1}}

\affiliation{
\aff{1}Department of Aerospace Engineering, Universidad Carlos III de Madrid, Av. de la Universidad, 30, Legan\'es, 28911, Madrid, Spain
\aff{2}Chair of Artificial Intelligence and Aerodynamics, School of Mechanical Engineering and Automation, Harbin Institute of Technology, 518055 Shenzhen, P.~R.~China
\aff{3}Department of Statistics, Universidad Carlos III de Madrid, Calle Madrid, 126, Getafe, 28903, Madrid, Spain
\aff{4} 
Guangdong Provincial Key Laboratory of Intelligent Morphing Mechanisms and Adaptive Robotics, Harbin Institute of Technology, 518055 Shenzhen, P.~R.~China
}

\begin{document}
\maketitle

\begin{abstract}\nolinenumbers
We propose a data-driven methodology to learn a low-dimensional manifold of controlled flows.
The starting point is resolving snapshot flow data for a representative ensemble of actuations.
Key enablers for the actuation manifold are isometric mapping as encoder, and a combination of a neural network and a $k$-nearest-neighbour interpolation as decoder.
This methodology is tested for the fluidic pinball,
a cluster of three parallel cylinders perpendicular to the oncoming uniform flow.
The centres of these cylinders  
are the vertices of an equilateral triangle pointing upstream. 
The flow is manipulated by constant rotation of the cylinders,
i.e.\ described by three actuation parameters.
The Reynolds number based on a cylinder diameter is chosen to be $30$.
The unforced flow yields statistically symmetric periodic shedding represented by a one-dimensional limit cycle.
The proposed methodology yields 
a five-dimensional manifold describing a wide range of dynamics with small representation error. 
Interestingly, the manifold coordinates automatically 
unveil physically meaningful parameters. 
Two of them describe the downstream periodic vortex shedding. 
The other three describe the near-field actuation,
i.e.\ the strength of boat-tailing, 
the Magnus effect and forward stagnation point. 
The manifold is shown to be a key enabler for control-oriented flow estimation.
\end{abstract}


\nolinenumbers
\section{Introduction}
\label{sec:intro}
In this study, we propose a data-driven manifold learner of the flows for a large range of operating conditions demonstrated by a low-dimensional manifold for the actuated fluidic pinball. A cornerstone of theoretical fluid mechanics is the low-dimensional representation of coherent structures. 
This representation is a basis for understanding dynamic modelling, 
sensor-based flow estimation, model-based control, and optimisation.

Many avenues of reduced-order representations have been proposed.
Leonardo da Vinci presented coherent structures and vortices artfully as sketches 
in a time when the Navier-Stokes equations were not known \citep{marusic2021leonardo}.
A quantitative pathway of low-dimensional modelling was started
by Hermann von \citet{Helmholtz1858} with his vortex laws.
Highlights are the \citet{karman1912mechanismus} model of the vortex street and derivation of feedback wake stabilization from the F\"oppl's vortex model by \citet{protas2004linear}.
Orr and Sommerfeld added a stability framework
culminating in Stuart's mean-field theory \citep{stuart1958non} to incorporate non-linear Reynolds stress effects.
The proper orthogonal decomposition (POD) has quickly become the major data-driven approach to compress snapshot data into a low-dimensional Galerkin expansion \citep{berkooz1993proper}.
Cluster-based modelling is an alternative data-driven flow compression, coarse-graining snapshot data into a small set of representative centroids \citep{kaiser2014cluster}.
All these low-dimensional flow representations 
are the kinematic prelude to dynamic models,
e.g.\ point vortex models, 
modal stability approaches \citep{theofilis2011global}, 
POD Galerkin models \citep{holmes2012turbulence}
and cluster-based network models \citep{fernex2021cluster}.

Data-driven manifold learners 
are a recent highly promising avenue of reduced-order representations.
A two-dimensional manifold may, for instance, accurately resolve
transient cylinder wakes:
the manifold dimension is a tiny fraction of the number of vortices, POD modes and clusters required for a similar resolution.
This manifold may be obtained by mean-field considerations \citep{noack2003hierarchy},
by dynamic features \citep{loiseau2018jfm},
by locally linear embedding \citep{Noack2023cambridge}
and by isometric mapping \citep[ISOMAP, ][]{FarzamnikISOMAPpinball2023}.
In all these approaches, the manifolds have been determined for a single operating condition. \citet{haller2023nonlinear} 
emphasised a key challenge,
namely, it remains \textit{``unclear if these manifolds are robust under parameter changes or under the addition of external, time-dependent forcing''.} In this study, we thus aim to identify a manifold representing fluid flows for a rich set of control actions by incorporating the actuation commands.

We choose the fluidic pinball as our benchmark problem
for the proposed manifold learning.
This configuration has been proposed by \citet{ishar2019metric}
as a geometrically simple two-dimensional configuration
with a rich dynamic complexity under cylinder actuation and Reynolds number change.
The transition scenario is described and modelled by \citet{Deng2020low}
comprising a sequence of bifurcations before passing to chaotic behaviour.
The feedback stabilisation achieved through cylinder rotation has been accomplished with many different approaches
\citep{Raibaudo2020pf, Cornejo2021stabilization, LiY2022jfm, wang2023cluster}.
\citet{FarzamnikISOMAPpinball2023} have demonstrated that the unforced dynamics 
can be described on a low-dimensional manifold. 

In this study, we illustrate the development of the control-oriented manifold
--- called an `actuation manifold' in the following ---
for the fluidic pinball with three steady cylinder rotations as independent inputs. ISOMAP is chosen for its superior compression capability and manifold interpretability over other techniques such as POD \citep{FarzamnikISOMAPpinball2023}.

The rest of the article is structured as follows. In \S \ref{sec:dataset} we introduce the dataset of the actuated fluidic pinball while in \S \ref{sec:method} we discuss the methodology employed to distill the actuation manifold and to use it for flow-estimation purposes. The results are presented and discussed in \S \ref{sec:results}, before highlighting the main conclusions of the work in \S \ref{sec:conclusions}.

\section{Flow control dataset}
\label{sec:dataset}

The dataset is based on two-dimensional incompressible uniform flow past three cylinders of equal diameter $D$. Their vertices form an equilateral triangle with one vertex pointing upstream and its median aligned with the streamwise direction. 
The origin of the Cartesian reference system is located in the midpoint between the two rightmost cylinders. The streamwise and crosswise directions are indicated with $x$ and $y$, respectively. The corresponding velocity components are indicated with $u$ and $v$. 
The computational domain $\Omega$ is bounded in $[-6,20]\times[-6,6]$, and the unstructured grid used has $4225$ triangles and $8633$ nodes. The snapshots are linearly interpolated on a structured grid with a spacing of $0.05$ in both $x$ and $y$ directions to maintain a format similar to the experimental data and for greater simplicity.

The Reynolds number ($Re$), which is based on $D$ and the incoming velocity $U_\infty$, is set to $30$ for all cases. A reference time scale $D/U_\infty$ is set as a convective unit (c.u.). The force coefficients $C_L$ and $C_D$ are obtained by normalising respectively lift and the drag forces with $\frac{1}{2} \rho U_\infty^2 D$, being $\rho$ the density of the fluid.

The control is achieved with independent cylinder rotations included in the vector $\boldsymbol{b} = (b_1, b_2, b_3)^T$, where $b_1, b_2$ and $b_3$ refer to the tangential speeds of the front, top and bottom cylinder, respectively. Here, '$T$' denotes the transpose operator. Positive actuation values correspond to counter-clockwise rotations. Combinations of rotational speed ranging from $-3$ to $3$ by steps of $1$ (i.e., $-3, -2, \ldots, 3$) have been used, for a total of 343 configurations, including the unforced case. Each of these simulation is run for $800$ c.u. to reach the steady state and snapshots with $1$ c.u. separation in the last $20$ c.u. for each steady state have been ¡sampled. This time is chosen to approximately cover two shedding cycles of the unforced case. This dataset therefore consists of $20$ post-transient snapshots for each of the $343$ combinations of actuations explored, for a total of $M = 6860$ snapshots.

The actuations are condensed into a three-parameter vector, denoted as $\boldsymbol{p} = (p_1, p_2, p_3)^T$, and referred to as Kiki parameters \citep{LinQ2021ms}. These three parameters describe the three mechanisms of boat-tailing ($p_1=(b_3-b_2)/2$), Magnus effect ($p_2=b_1+b_2+b_3$) and forward stagnation point control ($p_3 = b_1$). The Kiki parameters are introduced for an easier physical interpretation of the actuation commands.

\section{Methodology}
\label{sec:method}
\begin{figure}
    \centering
    \includegraphics{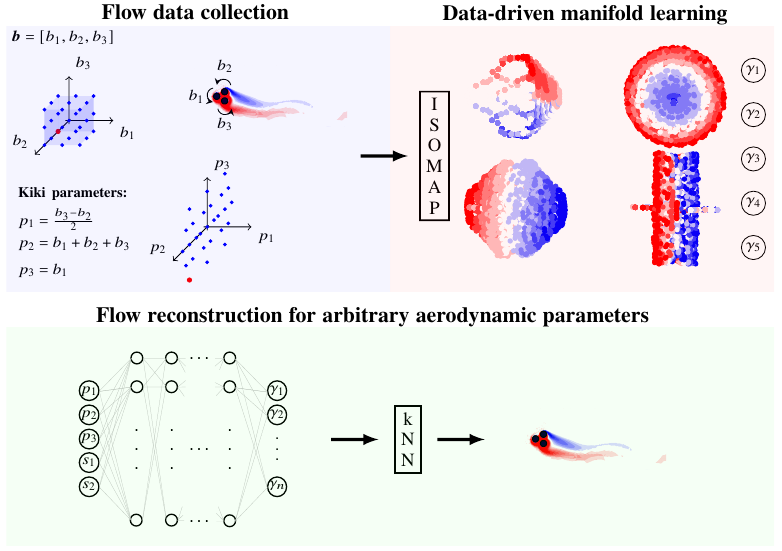}
    \caption{\justifying Illustration of the methodology for actuation-manifold learning and full-state estimation. The diagram highlights key steps, from flow data collection to data-driven actuation manifold discovery (upper section). A neural network, incorporating actuation parameters ($p_1$, $p_2$ and $p_3$) and sensor information ($s_1$ and $s_2$), determines ISOMAP coordinates ($\gamma_1, \gamma_2, \ldots, \gamma_n$) and a $k$NN decoder is used for the full-state flow reconstruction.}
    \label{fig:FIG1}
\end{figure}

A $n$-dimensional manifold is a topological space in which every point has a neighbourhood homeomorphic to Euclidean space $\mathbb{R}^n$. The approach proposed here involves employing an encoder-decoder strategy to acquire actuation manifolds, incorporating ISOMAP \citep{tenenbaum2000global}, neural networks and $k$-nearest neighbour ($k$NN) algorithms \citep{Fix1989KNN}. The collected dataset undergoes ISOMAP to discern a low-dimensional embedding. 
Then, the flow reconstruction procedure involves mapping actuation and sensor information to the manifold coordinates, followed by interpolation among the $k$-NNs to estimate the corresponding snapshot. An overview of this procedure is illustrated in figure~\ref{fig:FIG1} and detailed in the following.

If we consider our flow fields as vector functions $ \boldsymbol{u}(\boldsymbol{x})=(u(x,y),v(x,y))$ belonging to a Hilbert space, the inner product between two snapshots $\boldsymbol{u}_i$ and $\boldsymbol{u}_j$ is defined by:
\begin{equation}
\langle\boldsymbol{u}_i,\boldsymbol{u}_j\rangle= \iint_\Omega \boldsymbol{u}_i(x,y) \boldsymbol{{\cdot}} \boldsymbol{u}_j(x,y) \,dx\,dy
\end{equation}
where ``$\boldsymbol{\cdot}$'' refers to the scalar product in the two-dimensional vector space. Norms are canonically defined by $\|\bm{u}\| = \sqrt{\langle\boldsymbol{u},\boldsymbol{u}\rangle}$ and distances between snapshots are consistent with this norm. 

For our study, we have a collection of $M$ flow field snapshots. 
In the encoder procedure, the ISOMAP algorithm necessitates determining the square matrix $\mathsfbi{D}_{\rm G}\in\mathbb{R}^{M \times M}$ containing the geodesic distances among snapshots. Geodesic distances are approximated by selecting a set of neighbours for each snapshot to create a $k$NN graph. Within these neighbourhoods, snapshots are linked by paths whose lengths correspond to Hilbert-space distances. Geodesic distances between non-neighbours are then obtained as the shortest path through neighbouring snapshots in the $k$NN graph \citep{floyd1962algorithm}. At this stage, selecting the number of neighbours, which in the encoding part we denote by $k_e$, is crucial for creating the $k$NN graph and thus approximating the geodesic distance matrix.

Existing techniques are mostly empirical as in \citet{Samko2006OptimalK}. The method proposed by \citet{Samko2006OptimalK} was successfully used by \citet{FarzamnikISOMAPpinball2023}. However, due to the particularity of our dataset, as explained later in the article, standard empirical methods did not yield acceptable results. For this reason, we use the approach based on the Frobenius norm (denoted by $\|\cdot \|_F$) of the geodesic distance matrix $\mathsfbi{D}_{\rm G}$ \citep{Chao2005Optimalk} to determine $k_e$. Opting for a small $k_e$ can lead to disconnected regions and undefined distances within the dataset. Thus, there exists a minimum value of $k_e$ ensuring that all snapshots are linked in the $k$NN graph. Starting from this minimum, increasing $k_e$ results in more connections in the $k$NN graph, causing the Frobenius norm of the geodesic distance matrix to decrease monotonically. However, using an excessively large value may cause a short-circuiting within the $k$NN graph. Eventually, this yields an Euclidean representation, thus losing ISOMAP's capability to unfold nonlinear relationships within the dataset. Such short-circuiting is indicated by a sudden drop in $\|\boldsymbol{\mathsfbi{D}_{\rm G}}\|_F$ with increasing $k_e$. Thus, the minimum and the value at which short-circuiting first occurs define a suitable range of $k_e$ to be selected to approximate the geodesic distance matrix. The choice of $k_e$ within this range is generally based on how well the high-dimensional data are represented in the low-dimensional space. This is measured in terms of residual variance, as done by \citet{kouropteva2002selection}. 

After constructing the geodesic distance matrix $\mathsfbi{D}_{\rm G}$, multidimensional scaling \citep{torgerson1952multidimensional} is employed to construct the low-dimensional embedding $\mathbf\Gamma = (\tilde{\gamma}_{ij})_{1\le i\le M,\, 1\le j\le n}$, whose dimension is $n$. The $j$th column of the matrix $\mathbf\Gamma$ is the $j$th ISOMAP coordinate for all the $M$ flow field snapshots. We refer to the ISOMAP coordinates with $\gamma_{j}\,,j = 1,\ldots,n$. Consequently, $\tilde{\gamma}_{ij} = \gamma_j(i), ~i = 1,\ldots, M$.

More specifically, to obtain the low-dimensional embedding, we need to minimise the cost function:
\begin{equation}
    \|\mathbf\Gamma \mathbf\Gamma^T - \mathsfbi{B}\|_F^2
    \label{Eqn:MDSminim}
\end{equation}
where $\mathsfbi{B}$ is the double-centred squared-geodesic distance matrix $\mathsfbi{B} = -1/2 \mathsfbi{H}^T(\mathsfbi{D}_{\rm G} \odot \mathsfbi{D}_{\rm G})\mathsfbi{H}$. The centring matrix $\mathsfbi{H}$ is defined as $\mathsfbi{H} = \mathsfbi{I}_M - (1/M)\mathsfbi{1}_M$. Here $\mathsfbi{I}_M$ and $\mathsfbi{1}_M$ are the identity matrix and the matrix with all unit components of size $M\times M$, respectively. The symbol $\odot$ refers to the Hadamard product. The solution of the minimisation term in equation \eqref{Eqn:MDSminim} involves the eigenvalues and eigenvectors arising from the decomposition of $\mathsfbi{B}$, specifically $\mathsfbi{B} = \mathsfbi{V} \mathbf{\Lambda} \mathsfbi{V}^T$. For a given $n$, we select the $n$ largest positive eigenvalues, i.e.  $\mathbf{\Lambda}_n$, and the corresponding eigenvectors $\mathsfbi{V}_n$. The coordinates are obtained by scaling these eigenvectors by the square roots of their corresponding eigenvalues. Thus, the coordinates are given by $\mathbf{\Gamma} = \mathsfbi{V}_n \mathbf{\Lambda}_n^{1/2}$, where $\mathbf{\Lambda}_n^{1/2}$ is a diagonal matrix containing the square roots of the $n$ largest eigenvalues.

In this instance, following the methodology outlined by \cite{tenenbaum2000global}, the evaluation of data representation quality within the ISOMAP technique is conducted through the residual variance $R_v = 1 - R^2\left(\text{vec}\left(\mathsfbi{D}_{\rm G}\right), \text{vec}\left(\mathsfbi{D}_{\rm E}\right)\right)$. Here $R^2(\cdot,\cdot)$ denotes the squared correlation coefficient, 'vec' is the vectorisation operator and $\mathsfbi{D}_{\rm E}$ represents the matrix of the Euclidean distances between the low-dimensional representation of all the snapshots, retaining only the first $n$ ISOMAP coordinates. The residual variance indicates the ability of the low-dimensional embedding to reproduce the geodesic distances in the high-dimensional space. The dimension $n$ of the low-dimensional embedding is typically determined by identifying an elbow in the residual variance plot.

After the actuation manifold identification, the objective is to perform a flow reconstruction from the knowledge of a reduced number of sensors and actuation parameters. This process is carried out in two steps. First, a regression model is trained to identify the low-dimensional representation of a snapshot. Specifically, we employ a fully connected multi-layer perceptron (MLP) to map the actuation parameters and sensor information to the ISOMAP coordinates. It is possible to employ as network input either the actuation or the Kiki parameters. In the following, we employ the Kiki parameters since we deem them more elegant, although this does not substantially affect the results, being the Kiki parameters a linear combination of the actuation parameters. Second, a decoding procedure is conducted through linear interpolation among a fixed number (denoted by $k_d$) of nearest neighbours, following the methodology established by \cite{FarzamnikISOMAPpinball2023}. 
An alternative to this mapping is presented in appendix \ref{appA}, where a two-step $k$NN regression with distance-weighted averaging is employed. 
It is important to clarify that the decoding step is applicable only for interpolation cases, meaning for actuation cases that fall within the limits explored during the dataset generation, i.e., $|b_i| < 3,  ~ \forall i = 1,2,3$, with $|\cdot|$ being the absolute value.

\section{Results}
\label{sec:results}
\subsection{Identification and physical interpretation of the low-dimensional embedding}

\begin{figure}
    \centering
    \includegraphics{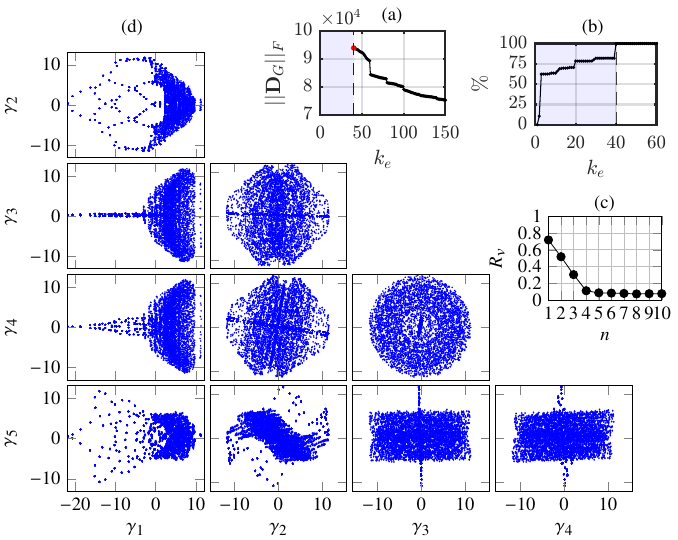}
    \caption{\justifying (\textit{a}) The Frobenius norm of the geodesic distance matrix plotted against the number of neighbours employed in Floyd's algorithm. (\textit{b}) The ratio between the connected snapshots and the total number of snapshots as $k_e$ increases. Results of the manifold obtained for $k_e = 40$ are presented in panels (c) and (d). The former illustrates the residual variance of the first $10$ ISOMAP coordinates, whereas the latter showcases all possible manifold sections identified by the first five coordinates.}
    \label{fig:Fig2}
\end{figure}

The results of the encoding step are represented in figure~\ref{fig:Fig2}. 
First, the neighbourhood size $k_e$ for the encoding must be determined. 
The Frobenius norm of the geodesic distance matrix $||\mathsfbi{D}_{\rm G}||_F$ as a function of $k_e$ is reported in figure~\ref{fig:Fig2}(a). 
A minimum $k_e = 40$ is required to ensure that all the neighbourhoods are connected. Figure~\ref{fig:Fig2}(b) reports the percentage of connected snapshots in the $k$NN graph. This percentage reaches $100\%$ for $k_e = 40$. Below this value, the manifold is divided into disjointed regions. By increasing $k_e$ starting from the minimum value, the Frobenius norm decreases due to improved connections between neighbourhoods. This also  raises the probability of short-circuits, which drastically reduces $||\mathsfbi{D}_{\rm G}||_F$. Indeed, the first short-circuiting is identified for $k_e = 60$. Choosing a value of $k_e$ within the range from the minimum value up to the point where the first significant drop in $||\mathsfbi{D}_{\rm G}||_F$ occurs does not result in substantial alterations of the geometry of the manifold or the physical interpretation of its coordinates. Therefore, in this paper, we select the minimum value, that is $k_e = 40$.

Then the residual variance is used to search for the true dimensionality of the dataset. Employing $n = 5$ coordinates leads to a residual variance of less than $10\%$, thus $n = 5$ dimensions are deemed sufficient to describe the manifold, as shown in figure~\ref{fig:Fig2}(c). Further increase in the number of dimensions provides only marginal changes in the residual variance.

The representation of the five-dimensional embedding is undertaken in figure~\ref{fig:Fig2}(d) in the form of two-dimensional projections on the $\gamma_i-\gamma_j$ planes with $i,j = 1,...,5$ and $i < j$. A first visual observation of the projections highlights interesting physical interpretations. The projection on the $\gamma_3-\gamma_4$ plane unveils a circular shape, suggesting that these two coordinates describe a periodic feature, i.e. the vortex shedding in the wake of the pinball. The fact that the circle is full suggests that different control actions result in an enhanced or attenuated vortex shedding. Each steady state describes a circle in the $\gamma_3 - \gamma_4$ plane. The radius of each circle explains the amplitude of vortex shedding, while the angular position provides information about the phase. Cases that do not exhibit vortex shedding collapse into points where $\gamma_3, \gamma_4 \approx 0$. This is evident from the observation of the projections on the planes $\gamma_1-\gamma_3$ and $\gamma_1-\gamma_4$, both returning a champagne coupe shape, suggesting that smaller values of $\gamma_1$ are a prerogative of the cases with limit cycle of smaller amplitude.

\begin{figure}
    \centering
    \includegraphics{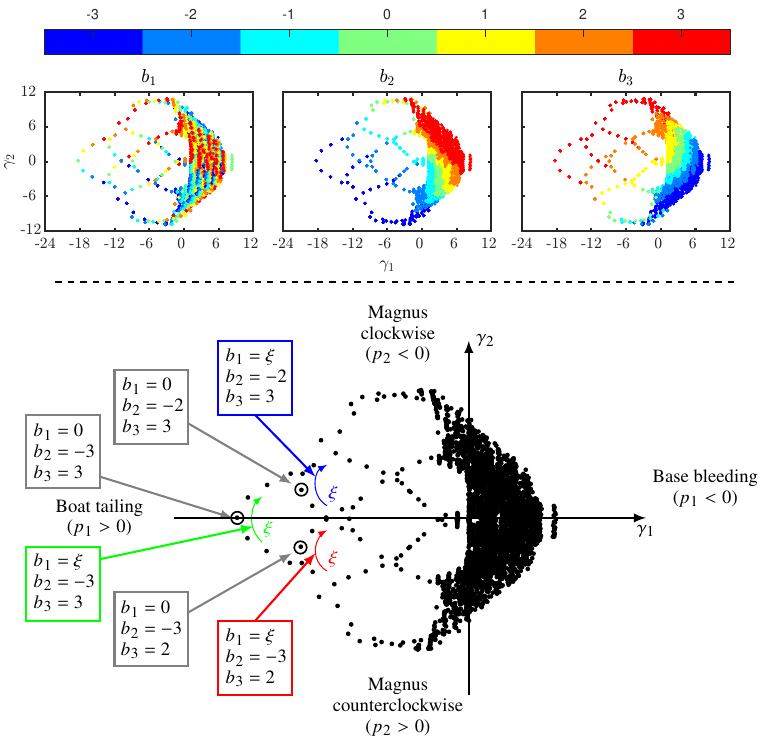}
    \caption{Three-dimensional projections of the manifold colour-coded with actuation parameters and forces coefficients. The first (boat-tailing) and second (Magnus) actuation parameters are plotted against lift and drag coefficients to understand physical control mechanisms.}
    \label{fig:Fig3}
\end{figure}

This feature also suggests that $\gamma_1$ is correlated with the boat tailing parameter $p_1$ and thus with the drag coefficient $C_D$, as visualised in figure~\ref{fig:Fig3}. This plot also shows a correlation between $\gamma_2$, the Magnus parameter $p_2$ and the lift coefficient $C_L$. The fifth coordinate of the low-dimensional embedding $\gamma_5$, on the other hand, appears to be correlated with the stagnation point parameter $p_3$ and partially explains the lift produced by the pinball. Intriguingly, all ISOMAP coordinates are physically meaningful and allow us to discover the three Kiki parameters without human input.

\begin{figure}
    \centering
    \includegraphics{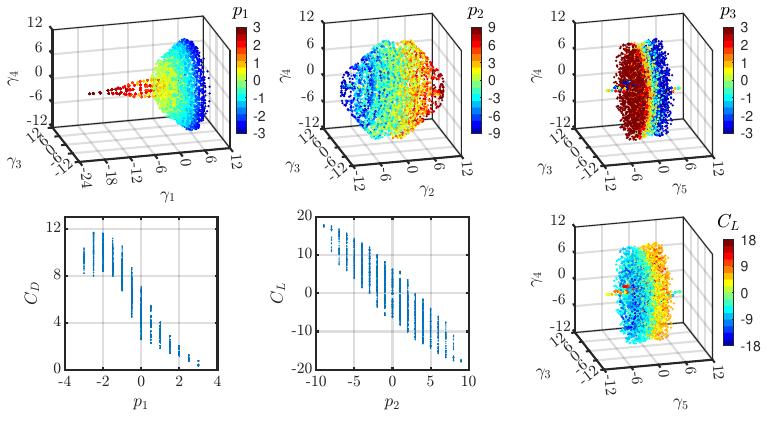}
    \caption{At the top, representation of the manifold section identified by the coordinates $\gamma_1$ and $\gamma_2$ colour coded with $b_1$ (left), $b_2$ (centre) and $b_3$ (right). At the bottom, a diagram explaining how the coordinate $\gamma_2$ provides indications of horizontal symmetry in the flow field.}
    \label{fig:Fig4}
\end{figure}

Another interesting physical interpretation arises observing the manifold section $\gamma_1-\gamma_2$, as shown in figure~\ref{fig:Fig4}. This provides insights into the horizontal symmetry of the data. In the figure, semicircles repeat, increasing in number as $\gamma_1$ increases. Within each of them, $b_1$ varies from $-3$ to $3$, whereas $b_2$ and $b_3$ remain fixed. The branches symmetrically positioned with respect to the axis $\gamma_2 = 0$ are associated with symmetric actuation $b_2$ and $b_3$.

The low-dimensional embedding, being the result of the eigenvector decomposition of the double-centred squared-geodesic distance matrix, is made of orthogonal vectors. 
While this might represent a disadvantage to autoencoders \citep{otto2022inadequacy}, this also allows projecting the snapshots on this basis and obtaining spatial modes that can be used to corroborate the physical interpretation of the manifold coordinates. The $j$th spatial mode $\boldsymbol{\phi}_j, \, j = 1,\ldots, n$ is a linear combination of the snapshots $\boldsymbol{u}_i$, i.e.~$({\sum_{i=1}^{M}\gamma_{ij}^2})^{1/2}\boldsymbol{\phi}_j= \sum_{i=1}^{M} \tilde{\gamma}_{ij} \boldsymbol{u}_i$.

\begin{figure}
    \centering
    \includegraphics{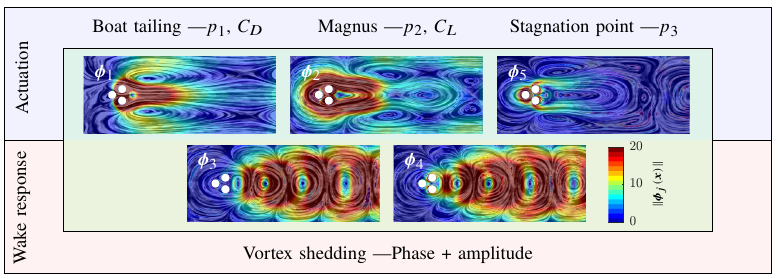}
    \caption{\justifying LIC representations of the normalised actuation modes. The shadowed contour represents the local velocity magnitude of the pseudomodes.}
    \label{fig:Fig5}
\end{figure}

The first five ISOMAP modes are visualised in figure~\ref{fig:Fig5} with a line integral convolution \citep[LIC;][]{forssell1995using} plot superimposed on a velocity magnitude contour plot. The visual observation of the modes confirms the interpretation of the physical meaning of the coordinates of the low-dimensional embedding.
Three modes, namely $\boldsymbol{\phi}_1$, $\boldsymbol{\phi}_2$ and $\boldsymbol{\phi}_5$ are representative of the actuation parameters.
The first ISOMAP mode is characterised by the presence of a jet (wake) downstream of the pinball due to boat-tailing (base bleeding). The second ISOMAP mode represents a circulating motion around the pinball, responsible for positive or negative lift, depending on the circulation direction. The fifth mode is characterised by a net circulation around the front cylinder, determining the position of the front cylinder stagnation point, if added to the mean field. 
The two spatial modes $\boldsymbol{\phi}_3$ and $\boldsymbol{\phi}_4$, instead, have the classical aspect of vortex shedding modes. Together, they describe the wake response to actuation parameters (far field), providing information on the intensity and phase of vortex shedding.

\subsection{Flow estimation}
For the identification of the low-dimensional representation of a snapshot, we employ a fully connected MLP. The mapping is done having as inputs the Kiki parameters and two sensors. The mapping also includes sensor information because each actuation vector is associated with a limit cycle, which in our case is described by $20$ post-transient snapshots. Therefore, the actuation vector provides comprehensive knowledge of the coordinates representing the near field ($\gamma_1$, $\gamma_2$ and $\gamma_5$). Far-field coordinates ($\gamma_3$ and $\gamma_4$) identification requires information about the phase of vortex shedding, which is provided by the sensors. Two different alternatives are proposed here. In the first, we utilise the lift coefficient and its one-quarter mean shedding period delay (we refer to this case with MLP$_1$). In the second, crosswise components of velocity are measured at two positions in the wake, specifically at points $\boldsymbol{x}_1 = (7,1.25)$ and $\boldsymbol{x}_2 = (10,1.25)$ (MLP$_2$). The sensors are located in the crosswise position at the edge of the shear layer of the unforced case, i.e. at the top of the upper cylinder. The streamwise location is selected such that the vortex shedding is already developed. The streamwise separation between the two sensors is such that the flight time between them is approximately one-quarter of the shedding period in the unforced case. This ensures that the sensor data collect relevant information for the flow reconstruction, even though a systematic position optimisation has not been performed. The characteristics of the employed networks are summarised in table \ref{tab:MLPparameters}. The training of the neural networks is performed with the dataset used to construct the data-driven manifold randomly removing all snapshots related to $20\%$ of the actuation cases and using the remaining snapshots for validation.

\begin{table} 
    \centering
    \begin{tabular}{m{0.2\textwidth} m{0.3\textwidth}m{0.2\textwidth}m{0.3\textwidth}}
     Input layer size     &  5 & Output layer size     &  5\\
     Number of samples    & {6860}    & Loss                 & {mean-square-error}\\
     Training set         & 80\%    & Batch size           & 2048\\ 
     Validation set       & 20\%   & Number of epochs     & Early stop on validation loss\\
     Test set             & {$22$ external cases} & Optimizer            & {Adam}\\
     I/O scaling & {Mean–std → $[0,1]$} & Activation function  & {$\tanh$}\\
     Hidden layers MLP$_1$ & ($70$,$70$,$70$,$70$) & Hidden layers MLP$_2$ & ($40$,$40$)\\
    \end{tabular}
    \caption{Structure of the neural networks used for latent coordinate identification.}
    \label{tab:MLPparameters}
\end{table}

To test the accuracy of flow estimation, we use $22$ additional simulations with randomly selected actuation, not present in the training nor the validation dataset. As done for the training dataset, the last $20$ c.u. are sampled every $1$ c.u. for a total of $440$ snapshots. The neural networks are used to identify 
the low-dimensional representation of the snapshots in the test dataset and then the $k$NN decoder is applied to reconstruct the full state.
In the decoding phase, we select $k_d = 220$. 

The value of $k_d$ was chosen by performing a parametric study. In particular, we selected the $k_d$ to minimise the reconstruction error of the snapshots in the validation dataset. The selected $k_d$ is considerably higher than the $k_e$. A reason for this result lies in the particular topology of the present dataset.
For some specific control actions, vortex shedding is completely suppressed, resulting in a stationary post-transient solution. The cases with no shedding have $20$ snapshots sampled in the post-transient identical to each other. Likewise, their representation in low-dimensional space is identical. When decoding using the $k$NN interpolation, for snapshots close to those without shedding, considering a low $k_d$ is equivalent to considering a small number of directions for the estimation of the gradients, which is not sufficient and results in a reconstruction with considerable error. To get a practical idea, considering $k_d<20$ would be equivalent to considering only one direction, which is insufficient for decoding. Such a dataset topology greatly complicates the choice of $k_d$, suggesting the need to identify a $k_d$ that is locally optimal. Such a choice, however, goes beyond the scope of the present research and is deferred to future work.

\begin{figure}
    \centering
    \includegraphics{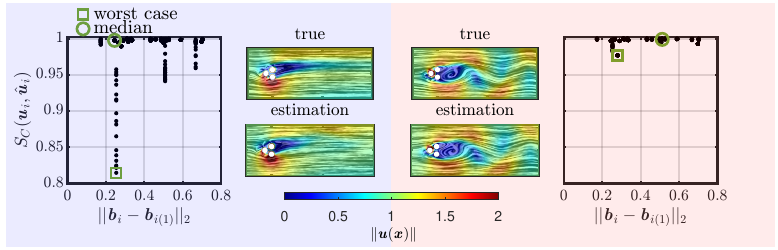}
    \caption{\justifying Cosine similarity between real and reconstructed snapshots for all $22$ actuation cases in the test dataset. Reconstruction error is plotted against the distance of each actuation case ($\boldsymbol{b}_i$) to its nearest neighbour in the training dataset $\boldsymbol{b}_{i(1)}$. In the center, a comparison between the true snapshot and its estimation is presented for the median error case. Figure presented with LIC and color-coded by velocity magnitude. The number of neighbours is fixed to $k_d = 220$. Results are shown for MLP$_1$ (left) and MLP$_2$ (right).}
    \label{fig:Fig6}
\end{figure}

The accuracy of the estimation between the true ($\boldsymbol{u}_i$) and estimated ($\hat{\boldsymbol{u}}_i$) snapshots is quantified in terms of cosine similarity $S_G\left(\boldsymbol{u}_i,\hat{\boldsymbol{u}}_i\right)=  \langle\boldsymbol{u}_i,\hat{\boldsymbol{u}}_i\rangle / \left(\|{\boldsymbol{u}}_i\|\|\hat{\boldsymbol{u}}_i\|\right)$. Results can be observed in figure~\ref{fig:Fig6}, the knowledge of the manifold enables a full state estimation with few sensors and minimal reconstruction error. When performing the flow reconstruction using MLP$_1$, the median cosine similarity is $0.9977$, corresponding to a median root-mean-square error (RMSE) of $0.0512$. Using MLP$_2$, instead, the median cosine similarity is $0.9990$ (RMSE = $0.0336$).

\section{Conclusions}
\label{sec:conclusions}
In this study, we have addressed a challenge of reduced-order modelling:
accurate low-dimensional representations for a large range of operating conditions. 
The flow data are compressed in a manifold 
using ISOMAP as encoder and $k$NN as decoder.
The methodology is demonstrated for the fluidic pinball 
as an established benchmark problem of modelling and control.
The cylinder rotations are used 
as three independent steady control parameters.
The starting point of the reduced-order modelling
is a data set with 20 statistically representative post-transient snapshots at $Re=30$ for $343$ sets of cylinder rotations covering uniformly a box with circumferential velocities between $-3$ and $3$.

The ISOMAP manifold describes all snapshots with $5$ latent variables 
and few per cent representation errors.
To the best of the authors' knowledge, 
this is the first demonstration of data-driven manifold learning 
for a flow under multiple input control.
Intriguingly, all latent variables are aligned with clear physical meanings. 
Three coordinates correspond to near-field actuation effects.
More precisely, these coordinates are strongly aligned with the Kiki parameters $\boldsymbol{p}$ introduced by \citet{LinQ2021ms},
describing (i) the level of boat-tailing (base bleeding) $p_1$ 
leading to drag reduction (increase),
(ii) the strength of the Magnus effect $p_2$ leading to a steady lift
and characterising (iii) the forward stagnation point $p_3$.
Two further coordinates correspond to the wake response to actuation,
i.e.\ the amplitude and phase of vortex shedding.
The distillation of physically meaningful parameters
from a fully automated manifold is surprising and inspiring (see figure~\ref{fig:FIG1}).

Our low-rank model is a key enabler for flow estimation with a minimum number of sensors. It is possible to estimate the full flow state with small reconstruction errors just by knowing the actuation parameters and employing two additional measurements, namely the aerodynamic forces or the flow velocity in two points of the wake.

We emphasise that the low-dimensional characterisation
of post-transient dynamics 
as a function of three actuations 
constitutes a modelling challenge 
and encourages numerous further applications,
e.g.\ vortex-induced vibration of a cylinder
at different spring stiffnesses and cylinder masses,
aerodynamic flutter of an elastic airfoil 
with different wing elasticities,
combustion instabilities for different reaction parameters,
just to name a few.
A low-dimensional manifold representation
can be utilised for understanding, 
estimation, prediction and model-based control.
In summary, the presented results can be expected 
to inspire a large range of applications 
and complementary manifold learning methods.

\backsection[Acknowledgements]{
We thank Qixin `Kiki' Lin for laying the groundwork of this study with her pioneering master thesis.}

\backsection[Funding]{This work is supported by the National Science Foundation of China (NSFC) through grants 12172109 and 12302293, by the Guangdong Basic and Applied Basic Research Foundation under grant 2022A1515011492, and by the Shenzhen Science and Technology Program under grants JCYJ20220531095605012, KJZD20230923115210021 and 29853MKCJ202300205, by the projects PITUFLOW-CM-UC3M, funded by the call “Programa de apoyo a la realización de proyectos interdisciplinares de I+D para jóvenes investigadores de la Universidad Carlos III de Madrid 2019-2020” under the frame of the Convenio Plurianual Comunidad de Madrid- Universidad Carlos III de Madrid and PREDATOR-CM-UC3M funded by the call ``Estímulo a la Investigación de Jóvenes Doctores/as'' within the frame of the Convenio Plurianual CM-UC3M and the V PRICIT (V Regional Plan for Scientific Research and Technological Innovation), by the project ARTURO, ref. PID2019-109717RB-I00/AEI/10.13039/501100011033, funded by the Spanish State Research Agency, by the project ACCREDITATION (Grant No TED2021-131453B-I00), funded by MCIN/AEI/ 10.13039/501100011033 and by the “European Union NextGenerationEU/PRTR”, by the funding under "Orden 3789/2022, del Vicepresidente, Consejero de Educación y Universidades, por la que se convocan ayudas para la contratación de personal investigador predoctoral en formación para el año 2022", by the European Research Council (ERC) under the European Union’s Horizon H2020 research and innovation programme (grant agreement No 949085) and by the project EXCALIBUR (Grant No PID2022-138314NB-I00), funded by MCIU/AEI/ 10.13039/501100011033 and by“ERDF A way of making Europe”. Funding for APC: Universidad Carlos III de Madrid (Agreement CRUE-Madroño 2024).}

\backsection[Declaration of interests] {The authors report no conflict of interest.}

\backsection[Data availability statement]{
The code used in this work is made available at: \href{https://github.com/Lmarra1/Actuation-manifold-from-snapshot-data.git}{https://github.com/Lmarra1/Actuation-manifold-from-snapshot-data.git}. The dataset is openly available in Zenodo, accessible
through the following link: \href{https://zenodo.org/doi/10.5281/zenodo.12802191}{https://zenodo.org/doi/10.5281/zenodo.12802191}
}

\backsection[Author ORCIDs]{\\
\orcidlink{0000-0001-9422-2808} L. Marra, 0000-0001-9422-2808;\\
\orcidlink{0000-0001-7499-7569} G. Y. Cornejo Maceda, 0000-0001-7499-7569;\\
\orcidlink{0000-0001-8537-9280} A. Meilán-Vila, 0000-0001-8537-9280;\\
\orcidlink{0000-0002-6610-7455} V. Guerrero,
0000-0002-6610-7455;\\
\orcidlink{0000-0001-5935-1962} B. R.\ Noack, 0000-0001-5935-1962;\\
\orcidlink{0000-0001-9025-1505} S. Discetti, 0000-0001-9025-1505;\\
\orcidlink{0000-0001-7342-4814} A. Ianiro,  0000-0001-7342-4814.}

\appendix
\section{Comparison between neural network and $k$NN performances in the decoding step}\label{appA}

In this appendix, we propose an alternative to the decoding as illustrated in figure~\ref{fig:FIG1}. 
In this approach, the decoding is carried out only utilising a $k$NN regression with weighted averaging with the inverse of distances between neighbours.
We transition from actuation parameters and sensor data to the low-dimensional representation of a snapshot using an initial $k$NN regression, with a number of neighbours indicated as $k_{d_1}$. Subsequently, we move from the manifold coordinates to the full state reconstruction using another $k$NN regression, with a number of neighbours indicated as $k_{d_2}$.

Consider the estimation of a snapshot $\boldsymbol{u}_i$ based on the actuation vector and sensor information in the vector $\boldsymbol{z}_i = (p_1, p_2, p_3, s_1, s_2)^T$. We estimate the low-dimensional representation of this snapshot, which we denoted as $\hat{\boldsymbol{y}}_i$, utilising the first $k$NN regression. Therefore, starting from $\boldsymbol{z}_i$, we consider its $k_{d_1}$-nearest vectors in the dataset, denoted as $\boldsymbol{z}_{i(1)}, \ldots, \boldsymbol{z}_{i(k_{d_1})}$, and their counterparts in the low-dimensional embedding $\boldsymbol{y}_{i(1)}, \ldots, \boldsymbol{y}_{i(k_{d_1})}$ to derive $\hat{\boldsymbol{y}}_i$ as follows:

\begin{equation}
    \hat{\boldsymbol{y}}_i = \frac{\sum_{j=1}^{k_{d_1}} \boldsymbol{y}_{i(j)}||\boldsymbol{z}_{i(j)} - \boldsymbol{z}_i||^{-1}}{\sum_{j=1}^{k_{d_1}} ||\boldsymbol{z}_{i(j)} - \boldsymbol{z}_i||^{-1}}
\end{equation}

Subsequently, the same procedure is applied to estimate the snapshot, denoted as $\hat{\boldsymbol{u}}_i$. We consider the $k_{d_2}$ nearest vectors to $\hat{\boldsymbol{y}}_i$, labelled as  $\hat{\boldsymbol{y}}_{i(1)}, \ldots, \hat{\boldsymbol{y}}_{i(k_{d_2})}$, and their counterparts in the high-dimensional space $\boldsymbol{u}_{i(1)}, \ldots, \boldsymbol{u}_{i(k_{d_2})}$ to obtain the estimate as follows:

\begin{equation}
    \hat{\boldsymbol{u}}_i = \frac{\sum_{j=1}^{k_{d_2}} \boldsymbol{u}_{i(j)}||\hat{\boldsymbol{y}}_{i(j)} - \hat{\boldsymbol{y}}_i||^{-1}}{\sum_{j=1}^{k_{d_2}} ||\hat{\boldsymbol{y}}_{i(j)} - \hat{\boldsymbol{y}}_i||^{-1}}
\end{equation} 

\begin{figure}
    \centering
    \includegraphics{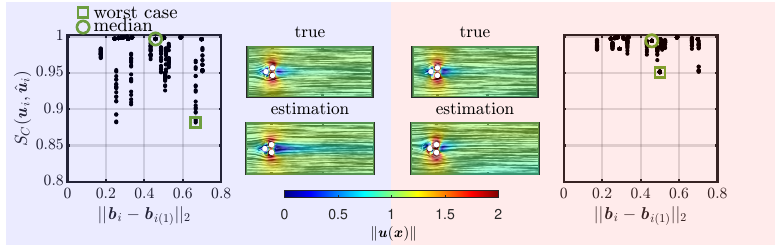}
    \caption{\justifying Cosine similarity between real and reconstructed snapshots for all $22$ actuation cases in the test dataset. Reconstruction error is plotted against the distance of each actuation case to its nearest neighbour in the training dataset. In the centre, a comparison between the true snapshot and its estimation is presented for the median error case. Figure presented with LIC and color-coded by velocity magnitude. Results are shown force for sensors (left) and velocity sensors (right).}
    \label{fig:Fig7}
\end{figure}

For the selection of $k_{d_1}$ and $k_{d_2}$ we have to solve the minimisation problems in \eqref{LpOCVm}:
\begin{equation}
  \left\{
  \begin{aligned}
   \underset{k_{d_1}}{argmin}\sum_{i=1}^{M}\left( \hat{\boldsymbol{y}}_i^{(-m), k_{d_1}} - \boldsymbol{y}_i \right)^2\\
    \underset{k_{d_2}}{argmin}\sum_{i=1}^{M}\left( \hat{\boldsymbol{u}}_i^{(-m), k_{d_2}} - \boldsymbol{u}_i \right)^2
  \end{aligned}
  \right.
  \label{LpOCVm}
\end{equation}

where $\hat{\boldsymbol{y}}_i^{(-m), k_{d_1}}$ and $\hat{\boldsymbol{u}}_i^{(-m), k_{d_2}}$ are the estimations of $\boldsymbol{y}_i$ and $\boldsymbol{u}_i$ using the $k$NN regression with $ k_{d_1}$ and $ k_{d_2}$ neighbours, respectively,
and using as the training dataset all snapshots except the $m = 20$ snapshots of the limit cycle to which the $i$th snapshot, being estimated, belongs. The procedure follows the same principles of the leave-one-out cross-validation \citep{Wand1994KernelSmoothing}, but instead of eliminating only one case for the estimation, all snapshots of the limit cycle corresponding to the execution are eliminated. The results of the snapshot reconstruction using $k$NN regression in the two steps of the decoding process are presented in figure~\ref{fig:Fig7}.

Table \ref{tab:KNNregression_k} shows the values of $k_{d_1}$ and  $k_{d_2}$ used for reconstructing the snapshots in the testing dataset for both sensing cases. In addition, the table also includes the reconstruction errors in terms of median cosine similarity and RMSE. The reconstruction performance is only slightly worse if compared to the case where a neural network followed by $k$NN interpolation is used.
\begin{table}
    \centering
    \begin{tabular}{ccccccc}
    & \multicolumn{4}{c}{ $k$NN+$k$NN regression} & \multicolumn{2}{c}{ MLP+$k$NN interpolation}\vspace{0.2cm}\\
                             & $k_{d_1}$ & $k_{d_2}$ & median $S_C$ & median RMSE & median $S_C$ & median RMSE  \vspace{0.1cm}\\
         Sensing forces      & $69$     & $7$ & $0.9970$ & $0.0573$ & $0.9977$ & $0.0512$\\
         Sensing velocities  & $66$     & $7$ & $0.9946$ & $0.0804$ & $0.9990$ & $0.0336$\\
    \end{tabular}
    \caption{Comparison of reconstruction errors in terms of cosine similarity and RMSE for both sensing cases using the two decoding methodologies: two-step $k$NN regression (left) and neural network and $k$NN interpolation (right). The table also includes the values of $k_{d_1}$ and $k_{d_2}$ used in the two-step $k$NN regression decoding method.} 
    \label{tab:KNNregression_k}
\end{table}


\end{document}